\documentstyle[12pt]{article}
\begin{document}
\def\strut{\rule[-.5cm]{0cm}{1cm}}
\def\dspace{\baselineskip = .30in}

\title{Heavy Top Quark From Fritzsch Mass Matrices\thanks{Supported in
part by Department of Energy Grant \#DE-FG02-91ER406267}}

\author{{\bf K.S. Babu}\\
\\and\\\\
{\bf Q. Shafi}\\
Bartol Research Institute\\
University of Delaware\\
Newark, DE 19716, USA}

\date{ }
\maketitle

\begin{abstract}
It is shown, contrary to common belief, that the Fritzsch ansatz for
the quark mass matrices admits a heavy top quark. With the ansatz
prescribed at the supersymmetric grand unified (GUT) scale, one finds
that the top quark may be as heavy as 145 GeV, provided that tan$\beta$
(the ratio
of the vacuum expectation values of the two higgs doublets)
$\gg 1$. Within a non-supersymmetric GUT framework with two
(one) light higgs doublets, the corresponding approximate upper bound
on the top mass is $120~ (90)$ GeV. Our results are based on a general
one--loop renormalization group analysis of the quark masses and mixing
angles and are readily applied to alternative mass matrix ans\"{a}tze.
\end{abstract}
\newpage

\dspace
The standard $SU(3) \times SU(2) \times U(1)$ gauge model predicts
the existence of a sixth (top) quark whose mass is known experimentally
to be heavier than 91 GeV.$^1$ Analysis based on electroweak
radiative corrections may
favor a moderately heavy top quark, say
below 160 GeV, with a central value possibly around the 120-140
GeV mark.$^2$ Other independent sources of information, such as
recovering the $b$ quark mass in the range $4.25 \pm 0.1$ GeV within
a supersymmetric grand unification framework (SUSY GUTs) with unified
third generation Yukawa couplings, also lead
to a top mass in the above range.$^3$

A simple approach for incorporating the observed hierarchy of quark
masses and their mixings was suggested a long time ago by
Fritzsch.$^4$ It prescribes a form for the
mass matrices which has a certain amount of predictive power.
Based on considerations of the $V_{cb}$ element of the
Cabibbo--Kobayashi--Maskawa (CKM) matrix,  it has been argued
that if the top quark mass turns out to be significantly
heavier than 90 GeV, then the simplest version of the Fritzsch ansatz
is excluded.$^{4,5}$

The main purpose of this note is to point out that the Fritzsch
ansatz actually permits the top quark to be much heavier,
provided it is considered within the framework of grand
unification involving more than one `light' higgs doublet.
Perhaps the most prominent example of this is provided by supersymmetric grand
unification. An important new parameter is $\tan\beta$, the well known
ratio of the vacuum expectation values of
the two higgs doublets that supersymmetry mandates. It turns out that
with $\tan\beta$ sufficiently greater than unity, the grand unified
supersymmetric version of the Fritzsch ansatz permits the top quark
to be as heavy as 145 GeV. A similar result holds within a non-supersymmetric
GUT framework provided that there are two `light' higgs doublets. In
this case the top quark mass can be as high as 120 GeV. For
completeness, we also present the results for a single higgs doublet and
confirm that the top mass cannot exceed by much 90 GeV.

Our analysis is based on the one--loop renormalization group evolution
of the quark masses and mixing angles.  The results are quite general and
can be used to
test alternative mass matrix ans\"{a}tze as well.  As an additional
application,
we derive lower limits on the top mass in the standard, two--higgs and
supersymmetric models assuming that the
relation $|V_{cb}^0|= \sqrt{m_c^0/m_t^0}$ holds at the unification scale.

The Fritzsh ansatz for the quark and charged lepton mass matrices
takes the following form at the GUT (or some other
appropriate superheavy) scale:

\begin{equation}
M_{u,d,\ell} = P_{u,d,\ell} \left( \begin{array}{ccc}
0 & a_{u,d,\ell} & 0\\
a_{u,d,\ell} & 0 & b_{u,d,\ell}\\
0 & b_{u,d,\ell} & c_{u,d,\ell}\end{array} \right) Q_{u,d,\ell}
\end{equation}

\noindent
where $P_{u,d,\ell}$ and $Q_{u,d,\ell}$ denote diagonal phase matrices
and $a,b,c$ are real
(positive) quantities.  Among the various phases contained in $P$ and $Q$, only
two are relevant for quark mixing,
we denote them by $\psi$ and $\phi$.  The ansatz
predicts the CKM matrix elements in terms of the quark
mass ratios and these two phase parameters.
In order to incorporate the fermion mass
hierarchy, one finds that $c \gg b \gg a$.
{}From the mass eigenvalues in the quark sector in particular,

\begin{equation}\begin{array}{lll}
c_u \simeq m^0_t, & b_u \simeq (m^0_cm^0_t)^\frac{1}{2}, & a_u \simeq
(m^0_um^0_c)^\frac{1}{2}\strut\\
c_d \simeq m^0_b, & b_d \simeq (m^0_sm^0_b)^\frac{1}{2}, & a_d \simeq
(m^0_dm^0_s)^\frac{1}{2}\end{array}~~.
\end{equation}

\noindent
The superscripts are to emphasize that the ansatz (1) is prescribed
at some superheavy scale.  [As a consequence of grand
unification the parameters $a_d, b_d, c_d$ of the down sector
typically are related to $a_\ell, b_\ell, c_\ell$ of the lepton
sector.]

Our main concern here is the following asymptotic relation which is a
consequence of eq. (1):

\begin{equation}
\mid V^0_{cb} \mid = \left| \sqrt{\frac{m^0_s}{m^0_b}} -
e^{i\phi} \sqrt{\frac{m^0_c}{m^0_t}} \right|~~.
\end{equation}

\noindent
The implications of (3) at low energies are evaluated by studying the
evolutions of the gauge and Yukawa couplings. In the absence of
Yukawa couplings that are comparable to or larger than the gauge
couplings, relation (3) would be essentially unrenormalized, and
would require that the top mass not exceed 90 GeV. However,
since the top quark mass is known to be greater than 91 GeV, and
indeed may be significantly larger than this, its Yukawa coupling
cannot be ignored. Moreover, models in which the up and down type
fermions obtain masses from couplings to distinct higgs doublets allow
for large intrinsic Yukawa couplings $h_b$ and $h_\tau$. We will
exploit, in particular, the fact that if $\tan\beta$ (the ratio of
the two doublet vev's which provide masses to up type and down type
fermions) is sufficiently large, then $h_b(h_\tau)$
can even exceed $h_t$! As a consequence, in contrast to the single
light higgs doublet case, relation (3) it turns out can lead to a low energy
prediction for $\mid V_{cb}\mid$ which is compatible with the
existence of a heavy top quark.

The Yukawa sector relevant for our discussion has the generic form

\begin{equation}
{\cal L}_Y = \bar{q}_L H_u \phi_u u_R + \bar{q}_L H_d \phi_d d_R +
\bar{\ell}_L H_\ell \phi_d e_R + h.c.
\end{equation}

\noindent
where
$H_u, H_d, H_\ell$ denote the $3 \times 3$ Yukawa coupling matrices
for the up quarks, down quarks and charged leptons.
Recall that in the standard model, $\phi_u \equiv \tilde{\phi}_d$.
The one loop evolution equations for the Yukawa matrices take the
form $(t \equiv \ell n (\mu/M_Z)$):$^6$
\begin{eqnarray}
16 \pi^2 {{dH_u}\over {dt}} & = & \left[ Tr (3 H_u H_u^{\dagger} +
3a H_d H_d^{\dagger}
+ a H_\ell H_\ell^{\dagger})\right.\nonumber \\
& & + \left. \frac{3}{2} (b H_u H_u^{\dagger} + c H_d H_d^{\dagger}) -
G_U \right] H_u\strut \nonumber \\
16 \pi^2 {{dH_d}\over {dt}} & = & \left[ Tr (3a H_u H_u^{\dagger} +
3H_d H_d^{\dagger} + H_\ell H_\ell^{\dagger}) \right. \nonumber \\
& & + \left. \frac{3}{2} (b H_d H_d^{\dagger} + c H_u H_u^{\dagger}) -
G_D \right]H_d\strut\\
16\pi^2 {{dH_\ell}\over {dt}} & = & \left[ Tr (3a H_u H_u^{\dagger} + 3 H_d
H_d^{\dagger} + H_\ell H_\ell^{\dagger}) \right.\nonumber \\
& & + \left. \frac{3}{2} b H_\ell H_\ell^{\dagger} -
G_E \right] H_\ell \nonumber ~~.
\end{eqnarray}

\noindent
For the three low energy models under discussion the coefficients
$a,b,c$ are given by

\begin{equation}\begin{array}{lcll}
(a,b,c) & = & (0,2,\frac{2}{3}) & {\rm MSSM\; (Minimal\; SUSY)}\strut\\
& = & (0,1,\frac{1}{3}) & {\rm Two\; higgs\; doublets\; (non-SUSY)}\strut\\
& = & (1,1,-1) & {\rm SM\; (Standard\; Model)}\end{array}
\end{equation}

\noindent
Also, the quantities $G_U, G_D$ and $G_E$ for the SUSY (non-SUSY)
case are respectively:

\begin{equation}\begin{array}{lcl}
G_U & = & \frac{13}{9} g^2_1 + 3g^2_2 + \frac{16}{3} g_3^2\strut\\
G_D & = & \frac{7}{9} g^2_1 + 3g^2_2 + \frac{16}{3} g^2_3\strut\\
G_E & = & 3g^2_1 + 3g^2_2~~;\end{array}~~~~{\rm (MSSM)}
\end{equation}
\vspace{.2in}

\begin{equation}\begin{array}{lcl}
G_U & = & \frac{17}{12} g^2_1 + \frac{9}{4} g^2_2 + 8 g^2_3\strut\\
G_D & = & \frac{5}{12} g^2_1 + \frac{9}{4} g^2_2 + 8 g^2_3\strut\\
G_E & = & \frac{15}{4} g^2_1 + \frac{9}{4} g^2_2~~.\end{array}
{}~~~~({\rm Two\; higgs\; ; SM})
\end{equation}

The gauge couplings $g_i$ (above) obey the standard one loop
renormalization group equations:

\begin{equation}
8 \pi^2 \frac{dg^2_i}{dt} = b_i g^4_i,~~~~ i = 1,2,3
\end{equation}

\noindent
where

\begin{equation}\begin{array}{lcll}
(b_1,b_2,b_3) & = & (11, 1, -3) & {\rm MSSM}\strut\\
& = & (7, 3, -7) & {\rm Two\; higgs}\strut\\
& = & (\frac{41}{6}, \frac{-19}{6}, -7) & {\rm SM}\end{array}
\end{equation}
Note that we are not using the $SU(5)$ normalization for the hypercharge
coupling $g_1$.
\noindent
{}From eq. (5), one can compute the evolution equations for the
eigenvalues of the Yukawa coupling matrices:$^{7,8}$

\begin{eqnarray}
16\pi^2 {{dg_i}\over {dt}} & = & g_i \left[ 3 \sum_{j=u,c,t} g^2_j + 3a
\sum_{\beta=d,s,b} g^2_\beta + a \sum_{b=e,\mu,\tau} g^2_b - G_U \right.
\nonumber \\
& & \left. + \frac{3}{2} bg^2_i + \frac{3}{2} c \sum_{\beta=d,s,b}
g^2_\beta \mid
V_{i\beta} \mid^2 \right]\strut \nonumber \\
16\pi^2 {{dg_\alpha}\over {dt}} & = & g_\alpha \left[ 3a \sum_{j=u,c,t}
g^2_j + 3 \sum_{\beta=d,s,b} g^2_\beta + \sum_{b=e,\mu,\tau} g^2_b -
G_D \right. \nonumber \\
& & \left. +  \frac{3}{2} b g^2_\alpha + \frac{3}{2} c \sum_{j=u,c,t} g^2_j
\mid
V_{j\alpha} \mid ^2 \right]\strut \nonumber \\
16\pi^2 {{dg_a}\over {dt}} & = & g_a \left[ 3a \sum_{j=u,c,t} g^2_j + 3 \right.
\sum_{\beta=d,s,b} g^2_\beta + \sum_{b=e,\mu,\tau} g^2_b - G_E \nonumber
\\
& & + \left. \frac{3}{2} b g^2_a \right]
\end{eqnarray}
where $i=(u,c,t),~\alpha=(d,s,b),~a=(e,\mu,\tau)$.

We will also need the evolution equations for the elements of the CKM
matrix:$^{7,8}$
\begin{eqnarray}
16\pi^2 {{d}\over {dt}} \mid V_{i\alpha} \mid^2 & = & 3c \left[
\sum_{j \neq i}\sum_{\beta=d,s,b} {{g^2_i + g^2_j}\over{g^2_i-g^2_j}
} g^2_\beta
Re \left( V_{i\beta} V^*_{j\beta} V_{j\alpha}
V^*_{i\alpha} \right)\strut \right. \nonumber \\
& & \left. + \sum_{\beta\neq \alpha}\sum_{j=u,c,t} {{g^2_\alpha +
g^2_\beta}\over {g^2_\alpha - g^2_\beta}} g^2_j Re \left( V^*_{j\beta}
V_{j\alpha} V_{i\beta} V^*_{i\alpha} \right)\right]~~.
\end{eqnarray}

The above expressions simplify considerably if we exploit the hierarchy
in the Yukawa couplings ($g_b \gg g_s \gg g_d$, etc) and in the CKM
matrix elements.  If
only the leading terms
are kept, one obtains the following approximate expressions for the
evolution of the various mass ratios and the mixing angles:
\begin{eqnarray}
16\pi^2 {{d}\over {dt}} \left( {{m_\alpha}\over {m_b}} \right) & = &
-\frac{3}{2} \left( {{m_\alpha}\over {m_b}} \right) \left( b g_b^2 + c
g^2_t), ~~~~\alpha = d,s \right.\strut \nonumber \\
16\pi^2 {{d}\over {dt}} \left( {{m_i}\over {m_t}} \right) & = &
-\frac{3}{2} \left( {{m_i}\over {m_t}} \right) (b g^2_t + c g^2_b),
{}~~~~i =
u,c\strut \nonumber \\
16\pi^2 {{d}\over {dt}} \left({{m_d}\over {m_s}}\right) & = & -\frac{3}{2}
\left({{m_d}\over {m_s}}\right) \left( b g^2_s + c g^2_c + c g^2_t \mid
V_{ts} \mid^2 \right)\strut \nonumber \\
16\pi^2 {{d}\over {dt}} \left( {{m_u}\over {m_c}} \right) & = & -
\frac{3}{2} \left({{m_u}\over {m_c}} \right) \left( b g^2_c + c g^2_s +
c g^2_b \mid V_{cb} \mid^2 \right)\strut \nonumber \\
16\pi^2 {{d}\over {dt}} \mid V_{i\alpha} \mid & = & - \frac{3}{2} c \mid
V_{i\alpha} \mid \left( g^2_t + g^2_b \right)\;~~~~ (i\alpha) = (ub), (cb),
(td), (ts)\strut \nonumber \\
16\pi^2 {{d}\over {dt}} \mid V_{us} \mid & = & - \frac{3}{2} c \mid
V_{us} \mid \left( g^2_c + g^2_s + g^2_t {{\mid V_{td}\mid^2 - \mid
V_{ub} \mid^2}\over {\mid V_{us} \mid^2}} \right)\strut \nonumber \\
16\pi^2 {{d}\over {dt}} \mid V_{cd} \mid & = & - \frac{3}{2} c \mid
V_{cd} \mid \left( g^2_c + g^2_s + g^2_b {{\mid V_{ub}\mid^2 -
\mid V_{td} \mid^2}\over {\mid V_{cd}\mid^2}}\right)~~.
\end{eqnarray}

\noindent
Several comments are in order:

\begin{enumerate}
\item The quantities $\mid V_{ub} \mid, \mid V_{cb} \mid, \mid V_{td}
\mid$ and $\mid V_{ts} \mid$ have identical evolutions.

\item The evolutions of $\mid V_{us} \mid$ and $\mid V_{cd} \mid$
involve second family Yukawa couplings and consequently are
negligible.

\item $m_d/m_b$ and $m_s/m_b$ have identical evolutions. Similarly
for $m_u/m_t$ and $m_c/m_t$.

\item The evolutions of $m_d/m_s$ and $m_u/m_c$ can be ignored.
\item The CP violating parameter $J$ is not independent, its evolution
can be obtained from the running of the square of $|V_{cb}|$, for
example.
\end{enumerate}

Using eq. (13) the asymptotic expression (3) can be recast in the form

\begin{equation}
\eta_{cb} \mid V_{cb} \mid = \left|\eta^\frac{1}{2}_{sb}
\sqrt{\frac{m_s}{m_b}} - e^{i\phi} \eta^\frac{1}{2}_{ct}
\sqrt{\frac{m_c}{m_t}}\right|
\end{equation}
where $|V_{cb}|$ stands for the CKM angle at the weak scale, and the
$\eta$'s are the respective running factors, $\eta_{cb}=|V_{cb}^0|
/|V_{cb}|$, $\eta_{sb} = (\frac{m_s^0}{m_b^0})/(\frac{m_s}{m_b})$, etc.
To compute these $\eta$ factors,
one has to solve numerically the
evolution equations in eq. (13).  The running
of $g_t,g_b,g_\tau$ are obtained from eq. (11), neglecting the light two
families.
As input parameters we take

\begin{equation}\begin{array}{lcl}
\alpha_1 (M_Z) & = & 0.01013\strut\\
\alpha_2 (M_Z) & = & 0.03322\strut\\
\alpha_3 (M_Z) & = & 0.115~~.\end{array}
\end{equation}

\noindent
For the superheavy scale $M_X$ above which the asymptotic relations
are valid we take

\begin{equation}\begin{array}{lcl}
M_X & = & 10^{16}\; {\rm GeV\; (SUSY)}\strut\\
& = & 10^{14}\; {\rm GeV\; (non-SUSY)}~~.\end{array}
\end{equation}

\noindent
For the running quark masses we use the values of ref. (9). In
particular, $m_s (1$ GeV) $= 175 \pm 55~ MeV, ~m_c (m_c) =1.27 \pm
0.05~ GeV$ and $m_b (m_b) = 4.25 \pm 0.1~ GeV$.

To get a feeling for the
effect running has on the prediction for $|V_{cb}|$, let us first define
the functions
\begin{equation}
f_{t,b} = {\rm exp}\left[{{1 \over {16 \pi^2}}{\int}^
{ln(M_X/M_Z)}_0 g_{t,b}^2(\tau) d\tau}\right]~~.
\end{equation}
Note that $f_{t,b}\ge 1$.  Let us first consider the supersymmetric
case.  For small tan$\beta$, $g_b$ can be
neglected since $g_b \ll g_t$.  The renormalized
prediction for $|V_{cb}|$ in terms of the mass ratios at the weak scale is
then
\begin{equation}
|V_{cb}| = \left|f_t^{\frac{1}{2}}\sqrt{{m_s}\over{m_b}} - e^{i \phi}
f_t^{-\frac{1}{2}}\sqrt{{m_c}\over{m_t}}\right|~~.
\end{equation}
Since $f_t \ge 1$, we see that the disagreement of $|V_{cb}|$
with observations is more prominent for small tan$\beta$.  This feature
persists as tan$\beta$ is increased, until it approaches $m_t/m_b$.
For tan$\beta = m_t/m_b$, $g_t \simeq g_b$ at all scales since both
couplings track the same evolution equations (except for small
corrections from the
hypercharge and the $\tau$ Yukawa couplings).  To a good
approximation, we see that the renormalized $|V_{cb}|$ obeys
\begin{equation}
|V_{cb}| = \left|\sqrt{{m_s}\over {m_b}}-e^{i\phi}\sqrt{{m_c}\over{m_t}}
\right|~~.
\end{equation}
That is, the relation (3) is essentially
unrenormalized and the top quark cannot be heavy.
This prompts us to consider values of
tan$\beta$ even larger than
$m_t/m_b$, in which case better agreement is possible.  Indeed, if we take the
extreme limit $g_b \gg g_t$, then in terms of $f_b$ defined in
eq. (17), we see that
\begin{equation}
|V_{cb}| = \left|f_b^{-\frac{1}{2}}\sqrt{{m_s}\over {m_b}} - e^{i \phi}
f_b^{\frac{1}{2}}\sqrt{{m_c}\over{m_t}}\right|~~.
\end{equation}
Note that the first term in eq. (20) has a reduction factor whereas the
second term is enhanced, precisely what one needs to bring the
prediction for $|V_{cb}|$ down and allow for a heavy top quark.
These features are indeed borne out by the actual numerical computation
to which we now turn.

In fig. 1a, we plot the behavior of the ratios $g_t(M_X)/g_t(M_Z)$,
$g_b(M_X)/g_b(M_Z)$ and $g_\tau(M_X)/g_\tau(M_Z)$ as  functions of
$m_t$ for tan$\beta=3$.  Throughout this paper, we shall mean by
$m_t$ the
running mass $m_t(m_t)$.  The pole mass is related to the running mass
via
\begin{equation}
m_t^{\rm pole} = m_t(m_t)\left[1+ {4 \over 3} {{\alpha_3} \over {\pi}}
\right]~~.
\end{equation}
Fig. 1b shows the dependence of the functions
$f(M_X)/f(M_Z)$ for $f=|V_{cb}|, |m_s/m_b|$ and $|m_c/m_t|$.  The
behavior of $|V_{ub}|, |V_{td}|$ and $|V_{ts}|$ are identical to that of
$|V_{cb}|$.  Similarly, $|m_d/m_b|$ runs as $|m_s/m_b|$ and $|m_u/m_t|$
as $|m_c/m_t|$.  The running factor for the CP parameter $J$ is obtained
from the square of $|V_{cb}|$.  Since tan$\beta$ is small, the evolution
of $|m_s/m_b|$ coincides with that of $|V_{cb}|$ which can be seen from
eq. (13).  In figures 2a and 2b, we plot the same quantities for the
case of tan$\beta=m_t/m_b$.  Note that the running of $g_t$
and $g_b$ are almost identical.  Finally, in figures 3a and 3b
we plot these functions for tan$\beta=60$.  This value of tan$\beta$
corresponds
to the infra--red fixed point solution for $g_b$ and $g_\tau$.

{}From figures 1--3, one can compute the running factors $\eta$'s
entering in (14).
It is clear from the discussions above as well as from
figures 1--3 that unless tan$\beta$ is larger than $m_t/m_b$, the
situation for $|V_{cb}|$ is worse than the case with no running.
However, for sufficiently
large tan$\beta$, $|V_{cb}|$ is in the experimentally
allowed range$^{10}$
of $0.034 \le |V_{cb}| \le 0.054$.  We give in Table 1 the
respective running factors for $|V_{cb}|, (m_s/m_b)$ and $(m_c/m_t)$ for
tan$\beta=60$ and also list the lowest allowed value of $|V_{cb}|$.
To arrive at the latter, we choose
$m_s(1~GeV) = 120~MeV$, $m_b(m_b) = 4.35~GeV$,
$m_c(m_c) = 1.32~GeV$ and set the phase $\phi=0$.  We use two--loop QCD
renormalization group equations to extrapolate the light quark masses
from low energies to $M_Z$.
There is
an upper limit of about 145 GeV on the top mass in this case coming from the
requirement that $g_b$ and $g_\tau$ should remain perturbatively small
in the entire range from $M_Z$ to $M_X$.
One sees that for all
values of $m_t$ up to 145 GeV, $|V_{cb}|$ is in the experimentally allowed
range.

$$\begin{array}{|c|c|c|c|c|}\hline
& & & &\\
m_t & \eta_{cb} & \eta_{sb} & \eta_{ct} & \mid V_{cb} \mid_{\rm min}\\
& & & &\\\hline
80 & 0.790 & 0.510 & 0.763 & 0.023 \\
90 & 0.781 & 0.499 & 0.746 & 0.029\\
100 & 0.770 & 0.485 & 0.727 & 0.034\\
110 & 0.757 & 0.467 & 0.703 & 0.038\\
120 & 0.740 & 0.445 & 0.675 & 0.042\\
130 & 0.718 & 0.415 & 0.639 & 0.045\\
140 & 0.684 & 0.371 & 0.591 & 0.046\\
145 & 0.660 & 0.338 & 0.559 & 0.045\\\hline\end{array}$$
\vskip.2in

\begin{center}
{\bf Table 1.} \\
The running factors for $|V_{cb}|, (m_s/m_b)$
and $(m_c/m_t)$ in the \\
supersymmetric model with tan$\beta=60$ as
functions of $m_t$. \\
In the
last column is listed the lowest allowed value of $|V_{cb}|$.
\end{center}
\vspace{.2in}

The precise values of $m_t$, $|V_{cb}|$ etc. will depend somewhat on the
input value of $\alpha_3(M_Z)$.  For larger values of $\alpha_3$, the
top quark can be heavier by a few GeV.  For e.g.,
with $\alpha_3(M_Z) = 0.12$, the top mass is raised
by about 5 GeV.  Most of the dependence arises from
changes in extrapolating
the light quark masses from 1 GeV to $M_Z$.

In figures 4,5 and 6, we plot the same quantities for the two higgs
model for tan$\beta=(3, m_t/m_b,70)$.
The behavior of the various functions
is identical to the case of SUSY, except that due
to the smaller beta function coefficients, the variation is not as
pronounced.  In Table 2, we list the running factors corresponding to
tan$\beta=70$ (the infra--red fixed point for $g_b$ and
$g_\tau$)
and conclude that $m_t$ as large as 120 GeV is allowed.

$$\begin{array}{|c|c|c|c|c|}\hline
& & & &\\
m_t & \eta_{cb} & \eta_{sb} & \eta_{ct} & \mid V_{cb} \mid_{\rm min}\\
& & & &\\\hline
80 & 0.887 & 0.708 & 0.875 & 0.034\\
90 & 0.885 & 0.705 & 0.869 & 0.040\\
100 & 0.882 & 0.701 & 0.862 & 0.046\\
110 & 0.878 & 0.697 & 0.854 & 0.050\\
120 & 0.874 & 0.692 & 0.844 & 0.054\\
130 & 0.870 & 0.686 & 0.834 & 0.058\\\hline\end{array}$$
\vspace{.2in}
\begin{center}
{\bf Table 2.}\\
Running factors and lowest allowed $|V_{cb}|$ in
the \\
two higgs model with tan$\beta=70$.
\end{center}
\vspace{.2in}

For completeness, we also display in
figures 7a and 7b the variation of the relevant quantities in the standard
model.  Table 3 shows the $\eta$ factors from which it is clear that
$m_t$ cannot much exceed 90 GeV.  This feature can also be understood
qualitatively in terms of the function $f_t$ of eq. (17).  The
renormalized relation for $|V_{cb}|$ in the standard model case is
\begin{equation}
|V_{cb}| = f_t^{-\frac{3}{4}}\left|\sqrt{{m_s}\over{m_b}} - e^{i\phi}
f_t^{-\frac{3}{2}}\sqrt{{m_c}\over {m_t}}\right|~~.
\end{equation}
Although there is an overall suppression factor, the second term also becomes
small and so the cancellation between the two terms becomes inefficient.

$$\begin{array}{|c|c|c|c|c|}\hline
& & & &\\
m_t & \eta_{cb} & \eta_{sb} & \eta_{ct} & \mid V_{cb} \mid_{\rm min}\\
& & & &\\\hline
80 & 1.020 & 1.020 & 0.981 & 0.049 \\
90 & 1.026 & 1.026 & 0.975 & 0.054\\
100 & 1.033 & 1.032 & 0.969 & 0.059\\
110 & 1.040 & 1.040 & 0.961 & 0.064\\
120 & 1.049 & 1.049 & 0.953 & 0.067\\\hline\end{array}$$
\vskip.2in
\begin{center}
{\bf Table 3.} \\
Running factors and $|V_{cb}|_{\rm min}$ for \\
the case of standard model.
\end{center}
\vspace{.2in}

Turning now to some other
predictions of the Fritzsch ansatz, the CKM elements $|V_{us}|$, $|V_{ub}|$
and $|V_{td}|$
have the asymptotic forms
\begin{eqnarray}
|V_{us}^0| & = & \left|\sqrt{{{m_d^0}\over {m_s^0}}} - e^{i \psi}
\sqrt{{{m_u^0}\over{m_c^0}}}\right| \nonumber \\
|V_{ub}^0| & = & \left|{{m_s^0}\over{m_b^0}} \sqrt{{{m_d^0}\over {m_b^0}}}
+ e^{i\psi}\sqrt{{{m_u^0}\over{m_c^0}}}
\left(\sqrt{{{m_s^0}\over{m_b^0}}}-e^{i \phi}\sqrt{{{m_c^0}
\over{m_t^0}}}\right)\right| \nonumber \\
|V_{td}^0| & = & \left|{{m_c^0}\over {m_t^0}}\sqrt{{{m_u^0}\over
{m_t^0}}} + e^{i \psi} \sqrt{{{m_d^0}\over {m_s^0}}}\left(
\sqrt{{{m_c^0}\over {m_t^0}}}- e^{i \phi}\sqrt{{{m_s^0}\over {m_b^0}}}
\right)\right|~~.
\end{eqnarray}
The first relation involves only the first two family masses and is
therefore essentially unrenormalized.  The phase $\psi$ should be near
$\pi/2$ for agreement with the Cabibbo angle.  From the second
relation in (23), one can easily write down the renormalized value of
$|V_{ub}|$.  It turns out that for tan$\beta=60$ and $m_t=130~GeV$ in the
SUSY model, the magnitude of the first term
is only about 15\% of the second term.  Given that the phase
$\psi \simeq \pi/2$ to a good approximation, the first term can be
neglected, which leads to the weak scale relation
\begin{equation}
{{|V_{ub}|}\over {|V_{cb}|}} \simeq \sqrt{{{m_u}\over {m_c}}} \simeq
0.06~~,
\end{equation}
in good agreement with the recent charmless $B$ decay data.
The same
conclusion also follows for the two higgs doublet model.  As for
$|V_{td}|$, again one finds that the first term is negligible compared
to the second term, so that at the weak scale we have
\begin{equation}
{{|V_{td}|}\over {|V_{cb}|}} \simeq \sqrt{{m_d}\over {m_s}} \simeq 0.2~~.
\end{equation}
Finally, the renormalized value of the
CP violating parameter $J$
at the weak scale is given by (for $\psi=\pi/2,~\phi=0$)
\begin{equation}
J \simeq \sqrt{{{m_d}\over {m_s}}}\sqrt{{{m_u}\over {m_c}}}|V_{cb}|^2
\simeq 3 \times 10^{-5}~,
\end{equation}
consistent with observations.

In the lepton sector, the Fritzsch ansatz can lead to two successful mass
predictions provided that $a_d=a_l,~c_d=c_l$ in eq. (1).  Such relations
arise naturally in GUT's if the elements $a,~c$ arise from a higgs
$\overline{\bf 5}$ of $SU(5)$ or a higgs {\bf 10} of $SO(10)$.
Note that $b_d$ and $b_l$ should be independent.  The two asymptotic
mass relations are
\begin{equation}
m_b^0-m_s^0+m_d^0 = m_\tau^0-m_\mu^0+m_e^0~;~~m_d^0m_s^0m_b^0 =
m_e^0m_\mu^0m_\tau^0~~.
\end{equation}
The first relation would lead to a prediction of $m_b(m_b) = 4.2~GeV$ in
the SUSY model with $m_t=130~GeV$ and tan$\beta=60$, which is in good
agreement with the value derived in ref. (9).  The second relation would
lead to $m_d(1~GeV) \simeq 7~MeV$ (if $m_s \simeq 140~MeV$),
also in good agreement with observations.

Since our analysis has been quite general, it is readily applicable to other
mass ans\"{a}tze as well.  One particularly interesting case which has
recently attracted a fair amount of attention is the asymptotic
relation$^{11,12,13}$
\begin{equation}
|V_{cb}^0| = \sqrt{{{m_c^0} \over {m_t^0}}}
\end{equation}
The renormalized relation in this case can be written down as
\begin{equation}
\eta_{cb}|V_{cb}| = \eta_{ct}^{\frac{1}{2}}\sqrt{{{m_c}\over{m_t}}}
\end{equation}
In Table 4, we list $\eta_{cb},~\eta_{ct}$ and the corresponding renormalized
values of $|V_{cb}|$ for the case of the standard model.  It is clear
that there is a lower limit of about 170 GeV on the top quark mass in
this case.

$$\begin{array}{|c|c|c|c|}\hline
& & &\\
m_t & \eta_{cb} & \eta_{ct} & \mid V_{cb} \mid\\
& & &\\\hline
100 & 1.033 & 0.969 & 0.081 \\
130 & 1.060 & 0.944 & 0.068 \\
160 & 1.104 & 0.906 & 0.058 \\
170 & 1.124 & 0.890 & 0.054 \\
180 & 1.149 & 0.870 & 0.051 \\
190 & 1.180 & 0.847 & 0.048 \\
200 & 1.221 & 0.819 & 0.044\\\hline\end{array}$$
\vskip.2in

\begin{center}
{\bf Table 4.}\\
Test of the asymptotic relation $|V_{cb}^0| =
\sqrt{m_c^0/m_t^0}$ in the \\
standard model.  $m_c(m_c) = 1.32~GeV$ is
used.
\end{center}
\vspace{.2in}

How does this asymptotic relation fare in the supersymmetric and two
higgs models?  In the SUSY model, the value of $|V_{cb}|$ at weak scale
can be written down as
\begin{equation}
|V_{cb}| = f_t^{-\frac{1}{2}}f_b^{\frac{1}{2}}\sqrt{{{m_c}\over {m_t}}}
{}~~.
\end{equation}
Since both $f_t$ and $f_b$ are greater than unity, it is clear that
smaller values of tan$\beta$ will give better agreement.  A similar
conclusion holds for the two higgs model as well.
In Tables 5
and 6, we list the $\eta$'s and values of $|V_{cb}|$ as functions of
$m_t$ in the SUSY model and in the two higgs model for tan$\beta=3$.  In
the SUSY model, the top quark mass should be close to the infra--red
fixed point value of about 185 GeV.  In the two higgs model, we see that
there is no acceptable solution with $m_t \le 200~GeV$ and
tan$\beta=3$.  Larger values of tan$\beta$ are not displayed, since the
agreement is worse.

$$\begin{array}{|c|c|c|c|}\hline
& & &\\
m_t & \eta_{cb} & \eta_{ct} & \mid V_{cb} \mid\\
& & &\\\hline
100 & 0.972 & 0.919 & 0.084 \\
130 & 0.946 & 0.847 & 0.072 \\
160 & 0.895 & 0.717 & 0.063 \\
170 & 0.862 & 0.641 & 0.060 \\
180 & 0.800 & 0.511 & 0.057 \\
185 & 0.708 & 0.356 & 0.052\\\hline\end{array}$$
\vskip.2in

\begin{center}
{\bf Table 5.}\\
Test of the asymptotic relation $|V_{cb}^0| =
\sqrt{m_c^0/m_t^0}$\\
in the SUSY model with tan$\beta=3$.
\end{center}
\vspace{.2in}

$$\begin{array}{|c|c|c|c|}\hline
& & &\\
m_t & \eta_{cb} & \eta_{ct} & \mid V_{cb} \mid\\
& & &\\\hline
100 & 0.988 & 0.965 & 0.084 \\
130 & 0.978 & 0.936 & 0.074 \\
160 & 0.963 & 0.892 & 0.066 \\
170 & 0.955 & 0.872 & 0.064 \\
180 & 0.947 & 0.848 & 0.061 \\
190 & 0.935 & 0.818 & 0.059 \\
200 & 0.921 & 0.780 & 0.057\\\hline\end{array}$$
\vskip.2in
\begin{center}
{\bf Table 6.}\\
Test of the asymptotic relation $|V_{cb}^0| =
\sqrt{m_c^0/m_t^0}$ \\
in the two higgs model with tan$\beta=3$.
\end{center}
\vspace{.2in}

To summarize, the Fritzsch ansatz for the quark mass matrices permits
the top quark to be in the mass range suggested by the recent
analysis of precision electroweak
data as well as by other independent estimates.
However, it requires that the parameter $\tan\beta$ be considerably
greater than unity, even exceeding $m_t/m_b$. It would be interesting
to see whether this can be reconciled
with the scenario of radiative electroweak breaking in the SUSY model
which usually
requires that $\tan\beta$ be $\stackrel{_<}{_\sim} m_t/m_b$.  The ansatz
also predicts values for the CKM matrix elements $|V_{us}|,~|V_{ub}|$,
$|V_{td}|$ and the CP violating parameter $J$ that are in agreement with
observations.
\vspace{.2in}
\newpage
\section*{References}
\begin{enumerate}
\item F. Abe et. al., CDF Collaboration, Phys. Rev. Lett. {\bf 68}, 447
(1992).
\item J. Ellis and G. Fogli, Phys. Lett. {\bf B 249}, 543 (1990);\\
P. Langacker and M. Luo, Phys. Rev. {\bf D 44}, 817 (1991).
\item B. Anantanarayanan, G. Lazarides and Q. Shafi, Phys. Rev.
{\bf D 44}, 1613 (1991) and Proc. of the PASCOS Symposium, March (1991).
\item H. Fritzsch, Phys. Lett. {\bf B 70}, 436 (1977) and Nucl. Phys.
{\bf B 155}, 189 (1979).
\item F. Gilman and Y. Nir, Annu. Rev. Nucl. Part. Sci. {\bf 40}, 213
(1990);\\
E. Ma, Phys. Rev. {\bf D 43}, R2761 (1991).
\item T.P. Cheng, E. Eichten and L-F. Li, Phys. Rev. {\bf D 9}, 225 (1974);
\\
M. Machacek and M. Vaughn, Nucl. Phys. {\bf B 236}, 221 (1984).
\item E. Ma and S. Pakvasa, Phys. Lett. {\bf B 86}, 43 (1979) and Phys.
Rev. {\bf D 20}, 2899 (1979).
\item K.S. Babu, Z. Phys. {\bf C 35}, 69 (1987);\\
K. Sakai, Z. Phys. {\bf C 32}, 149 91986);\\
B. Grzadkowski, M. Lindner and S. Theisen, Phys. Lett. {\bf B 198}, 64
(1987);\\
M. Olechowski and S. Pokorski, Phys. Lett. {\bf B 257}, 388 (1991).
\item J. Gasser and H. Leutwyler, Phys. Rept. {\bf 87}, 77 (1982).
\item Particle Data Tables, Phys. Rev. {\bf D 45} (1992).
\item J. Harvey, P. Ramond and D. Reiss, Nucl. Phys. {\bf B 199}, 223
(1982).
\item E. Freire, G. Lazarides and Q. Shafi, Mod. Phys. Lett. {\bf A 5}, 2453
(1990).
\item S. Dimopoulos, L. Hall and S. Raby, Phys. Rev. Lett. {\bf 68},
1984 (1992);\\
G. Anderson, S. Raby, S. Dimopoulos and L. Hall, Ohio State Preprint,
September (1992);\\
H. Arason et. al., Florida Preprint UFIFT-HEP-92-8;\\
V. Barger, M. Berger and P. Ohmann, Wisconsin Preprint, September
(1992).
\end{enumerate}
\newpage
\section*{Figure Captions}

Fig. 1a:  Plot of $g_t(M_X)/g_t(M_Z)$ (solid), $g_b(M_X)/g_b(M_Z)$
(dashed) and \\
\hspace*{.6in} $g_\tau(M_X)/g_\tau(M_Z)$ (dot-dash)
as functions of $m_t(m_t)$ in the \\
\hspace*{.6in} supersymmetric model
with tan$\beta=3$.  \\
Fig. 1b:  The running factors $f(M_X)/f(M_Z)$ for $f=|V_{cb}|$ (solid),
$|m_s/m_b|$ \\
\hspace*{.6in} (dashed) and $|m_c/m_t|$ (dot--dash) versus $m_t$.  The
running factors \\
\hspace*{.6in} for $|V_{ub}|,~|V_{td}|$ and $|V_{ts}|$ are identical to
that of $|V_{cb}|$.  Similarly, \\
\hspace*{.6in} $|m_d/m_b|$ runs as $|m_s/m_b|$ and
$|m_u/m_t|$ as $|m_c/m_t|$.  \\
Fig. 2a:  Same as in fig. 1a, except that tan$\beta=m_t/m_b$.  \\
Fig. 2b:  Same as in fig. 1b, except that tan$\beta=m_t/m_b$.  \\
Fig. 3a:  Same as in fig. 1a, except that tan$\beta=60$.  \\
Fig. 3b:  Same as in fig. 1b, except that tan$\beta=60$.  \\
Fig. 4a:  The running factors for the two higgs (non--SUSY) model with \\
\hspace*{.6in} tan$\beta=3$.  Notation same as in fig. 1a. \\
Fig. 4b:  Running factors for the two higgs (non--SUSY) model with
tan$\beta=3$,\\
\hspace*{.6in} notation same as in fig. 1b. \\
Fig. 5a:  Same as in fig. 4a, except that tan$\beta=m_t/m_b$.  \\
Fig. 5b:  Same as in fig. 4b, except that tan$\beta=m_t/m_b$.  \\
Fig. 6a:  Same as in fig. 4a, except that tan$\beta=70$.  \\
Fig. 6b:  Same as in fig. 4b, except that tan$\beta=70$.  \\
Fig. 7a:  The running factors for the minimal standard model in the same \\
\hspace*{.6in} notation as fig. 1a. \\
Fig. 7b:  Running of the mixing angles and mass ratios in the standard \\
\hspace*{.6in} model, notation same as in fig. 1b.

\end{document}